\documentclass[]{iopart}
\usepackage{amsbsy}
\usepackage{latexsym}
\usepackage{graphicx}
\usepackage{url}
\usepackage{acronym}

\begin{document}

\acrodef{LIGO}{Laser Interferometer Gravitational-wave Observatory}
\acrodef{aLIGO}{advanced LIGO}
\acrodef{LCGT}{Large Cryogenic Gravitational Wave Telescope}
\acrodef{SNR}{signal-to-noise ratio}
\acrodef{BNS}{binary neutron star}

\title[Localization with advanced GW network]{Source localization with an
advanced gravitational wave detector network}

\author{Stephen Fairhurst}
\address{Cardiff School of Physics and Astronomy,
Cardiff University, Queens Buildings, The Parade, Cardiff. CF24 3AA
}
\eads{\mailto{Stephen.Fairhurst@astro.cf.ac.uk}}

%=====================
\begin{abstract} 
We derive an expression for the accuracy with which sources can be
localized using a network of gravitational wave detectors.  The result
is obtained via triangulation, using timing accuracies at each detector
and is applicable to a network with any number of detectors.  We use
this result to investigate the ability of advanced gravitational wave
detector networks to accurately localize signals from compact binary
coalescences.  We demonstrate that additional detectors can
significantly improve localization results and illustrate our findings
with networks comprised of the advanced \ac{LIGO}, advanced Virgo and
\ac{LCGT}.  In addition, we evaluate the benefits of relocating one of
the advanced \ac{LIGO} detectors to Australia.
\end{abstract}

%=====================
\section{Introduction}
\label{sec:intro}
%=====================

The era of advanced gravitational wave detectors \cite{Harry:2010, lcgt,
adV} is rapidly approaching, and with it the prospect of the regular
observation of gravitational wave signals.  To fully exploit these
gravitational wave observations, it will be critically important to
correlate them whenever possible with electromagnetic counterparts.  For
example, observation of gravitational waves coincident with Gamma Ray
Bursts will aid in determining the nature of the progenitor
\cite{nakar07}; the gravitational wave signal from core-collapse
supernovae may carry information about the supernova engine \cite{Ott};
gravitational and electromagnetic observations of binary coalescence
will provide independent measures of the distance and redshift to the
source and consequently enable precision tests of cosmology
\cite{Schutz:1986gp}.  One method of performing multi-messenger
observations will be to accurately and rapidly localize the source
through gravitational wave observations and then use electromagnetic
observatories to follow up on the event.  Indeed, an ambitious project
to follow up gravitational wave candidates in a host of electromagnetic
telescopes is already underway \cite{Kanner:2008}.  A single
gravitational wave detector gives virtually no directional information
for a short duration signal.  Consequently, localization requires a
network of detectors and the primary tool for localization is
triangulation between a signal observed at several sites
\cite{Fairhurst:2009tc}.  

A number of electromagnetic telescopes are being designed, built and
operated specifically to search for transient phenomena.  These include
Palomar Transient Factory \cite{Rau:2009yx}, Pan Starrs
\cite{panstarrs}, SkyMapper \cite{skymapper}, LOFAR \cite{lofar} and
many more.  These observatories will have a field of view of around ten
square degrees.  There are plausible astrophysical mechanisms that
produce both gravitational waves and electromagnetic signals in various
frequencies, for example gamma ray, x-ray, optical and radio (see, e.g.
\cite{nakar07, Metzger:2010sy, Predoi:2009af, ChassandeMottin:2010zh}).
Thus, providing a good estimate of the localization capability of the
gravitational wave network will help inform the follow-up searches which
are undertaken.  In addition, if the source can be localized to a region
which contains only one, or possibly a few, galaxies then a narrow field
telescope could be used to image the galaxy in question
\cite{Nuttall:2010nk}.    

In this paper, we investigate the accuracy with which gravitational wave
sources can be localized using a network of advanced detectors.  A
network comprising more detectors, particularly if they are at widely
separated sites, gives better localization.  Thus, the recent
announcement of funding for the construction of the Japanese 
\ac{LCGT} \cite{lcgt} is
particularly welcome.  This detector will augment the \ac{aLIGO}
detectors \cite{Harry:2010} to be installed at sites at Hanford, WA and
Livingston, LA in the USA and the advanced Virgo \cite{adV} detector to
be installed in Cascina, Italy and provide a four site network.
Furthermore, there is significant interest in installing one of the
three advanced LIGO detectors at the Gingin site in Australia
\cite{LIGOSouth, AdrianCho08272010}, providing a realistic prospect for
a five site network of advanced gravitational wave detectors.  

The layout of the paper is as follows.  In Section
\ref{sec:triangulation} we extend the framework presented in
\cite{Fairhurst:2009tc} to provide a simple, geometrical expression for
the localization region obtained with any number of detectors.  Then, in
Section \ref{sec:adv_networks}, we examine the capacity of the various
networks to localize sources.  The formalism is equally applicable to
any transient source.  However, for simplicity we present a sample of
results for \ac{BNS} mergers.  As noted in
\cite{Fairhurst:2009tc}, the results for both neutron star--black hole
and binary black hole mergers with a total mass below around $10
M_{\odot}$ will be comparable.

%=====================
\section{Sky Localization from Triangulation}
\label{sec:triangulation}
%=====================

For the purposes of this paper, we assume that localization is achieved
entirely by triangulation of the source by the network of gravitational
wave detectors.  In reality, there is additional information available
in the waveform; for instance the fact that the gravitational waveform
consists of only two polarizations can be used to coherently combine the
data streams and improve localization \cite{GurselTinto, Veitch:2009hd,
Christensen}.  Indeed, even the non-observation of an event in one
detector in the network can be used to assist with localization.
Generically, these effects will serve to improve the localization of
events.  In addition, there will be numerous systematic uncertainties in
estimating the localization of signals; for example imperfect
calibration of the detectors and uncertainties in the emitted
gravitational waveform.  We will not investigate these in great detail,
but merely note that systematic errors are expected to be smaller than
statistical uncertainties for all but the loudest sources
\cite{Fairhurst:2009tc}.

%=====================
\subsection{Timing}
\label{ssec:timing}

The accuracy with which the time%
\footnote{For an extended signal such as a binary coalescence, this
should be regarded as a fiducial time, for example the time at which the
frequency of the signal passes through 100 Hz.}
 a gravitational wave signal passed a given detector can be determined
is given by \cite{Fairhurst:2009tc}%
\footnote{A similar expression was obtained previously in
\cite{Schutz:1992jk}.}
\begin{equation}\label{eq:sigma_t}
  \sigma_{t} = \frac{1}{2 \pi \rho \sigma_{f}} \, .
\end{equation}
Timing accuracy is inversely proportional to both the \ac{SNR} $\rho$ and
effective bandwidth $\sigma_f$ of the source, defined as 
\begin{eqnarray}
  \rho^{2} &=& 4 \int_{0}^{\infty} \frac{| h(f) |^{2}}{S(f)} df \, , 
    \nonumber \\
  \sigma_{f}^{2} &=& 
    \left(\frac{4}{\rho^{2}} \int_{0}^{\infty} df 
    \frac{f^2 |h(f)|^2}{S(f)}\right) - 
    \left(\frac{4}{\rho^{2}} \int_{0}^{\infty} df 
    \frac{f |h(f)|^2}{S(f)} \right)^{2} \, ,
\end{eqnarray}
where $S(f)$ denotes the one sided noise power spectral density of the
detector.  The approximations used to obtain this formula break down at
low \ac{SNR}, where second order effects become important
\cite{Vitale:2010mr}.  At an \ac{SNR} of 8, which we consider in several
later examples, this is about a 10\% effect.

\begin{table}[t]
\center
\begin{tabular}{| l | c | c | c | c | c| }
\hline \hline
Detector  & Horizon & Range & $\overline{f}$  & $\sigma_{f}$ & 
  $\sigma_{t}$ at $\rho=8$ \\ 
 & (Mpc) & (Mpc) & (Hz) & (Hz) & (ms) \\
\hline
\hline
\ac{aLIGO} NOSRM \cite{aLIGO}& 360 & 160 & 65 & 43 & 0.46 \\
\ac{aLIGO} BNS \cite{aLIGO}& 490 & 215 & 110 & 106 & 0.19 \\
Advanced Virgo \cite{adV}& 350 & 155 & 120 & 100 & 0.20 \\
\ac{LCGT} \cite{lcgt} & 365 & 160 & 100 & 88 & 0.22 \\
\hline
\hline
\end{tabular}
\caption{Currently projected advanced detector sensitivity for \ac{BNS}
systems.  The sensitivity of the detector is encoded in the \textit{\ac{BNS}
horizon} (distance at which an optimally oriented and located signal
gives an \ac{SNR} of 8) and the \textit{\ac{BNS} range} (the volume and
orientation averaged distance at which a \ac{BNS} gives \ac{SNR} 8).  The
\textit{horizon} is a factor of 2.26 larger than the \textit{range}.
The mean frequency $\bar{f}$ and frequency bandwidth give the frequency
range in which the detector is sensitive to the signal.  The frequency
bandwidth $\sigma_{f}$ determines the timing width $\sigma_{t}$ at a
given \ac{SNR} through (\ref{eq:sigma_t}).  The two sets of advanced LIGO
numbers correspond to no signal recycling (NOSRM) and \ac{BNS} optimized
(BNS) configurations.}
\label{tab:bns_time}
\end{table}

In Table \ref{tab:bns_time}, we provide the currently projected
sensitivities, frequency bandwidths and timing accuracies for the
various advanced gravitational wave detectors.  Numerous configurations
have been proposed for the advanced detectors \cite{Harry:2010}, and it
is likely that several will be used over the lifetime of the detectors.
To illustrate the differences, we consider two \ac{aLIGO} configurations, one
with no signal recycling mirror (NOSRM) which is likely to be an early
configuration, and one optimized for \ac{BNS} detection (BNS) which may be
used in later science runs.  The anticipated noise curves represent the
incoherent sum of the principal noise sources as currently understood.
There will be, in addition, technical noise sources.  The sensitivities
in Table \ref{tab:bns_time} are not the guaranteed performance of advanced
detectors, but a good guide to the anticipated sensitivity.

For the most part, advanced detectors will be sensitive to an optimally
oriented and located \ac{BNS} to a \textit{horizon} distance of around
360 Mpc, although the \ac{BNS} optimized \ac{aLIGO} configuration
provides a \textit{horizon} of almost 500 Mpc.  The frequency bandwidth
of a \ac{BNS} signal in the detectors will be around $100 \mathrm{Hz}$,
leading to a timing accuracy of $0.2 \mathrm{ms}$ at an \ac{SNR} of 8.
The \ac{aLIGO} NOSRM configuration has a noise curve which rises more
sharply at high frequency. Consequently, it has a significantly smaller
bandwidth and a timing accuracy of almost $0.5 \mathrm{ms}$ at \ac{SNR}
8, more than a factor of 2 larger than the other configurations listed.

For the remainder of this paper, we make the simplifying assumption that
all the advanced detectors have the same sensitivity and bandwidth.  Our
standardized advanced detector will have a \textit{\ac{BNS} horizon} of
360 Mpc (corresponding to a sky and orientation averaged
\textit{\ac{BNS} range} of 160 Mpc at SNR 8) and a bandwidth of 100 Hz.
It is relatively straightforward to scale the results to other parameter
choices using Equation (\ref{eq:sigma_t}).

%=====================
\subsection{Localization}
\label{ssec:localization}

The measured time of arrival of a signal in a network of detectors can
be used to reconstruct the source location.  However, consideration of
the observed amplitudes and phases of the signal at all detectors in the
network is critical for extracting the full set of parameters, including
the distance to the binary and its orientation
\cite{PaiDhurandharBose2001, Veitch:2009hd}.  For our purposes a
gravitational wave signal is described by the location $\mathbf{R}$ of
the source and the time $T_{o}$ at which the signal passes through the
center of the earth.  Since our primary focus is localization, we take
$\mathbf{R}$ to be a unit vector describing only the position of the
source and not its distance.

The time at which the signal passes through detector $i$ is given by
\begin{equation}\label{eq:T_i}
  T_{i} = T_{o} - \mathbf{R} \cdot \mathbf{d}_{i} \, ,
\end{equation}
where $\mathbf{d}_{i}$ encodes the separation between detector $i$ and
the center of the earth (expressed in seconds).  The distribution of the
measured arrival times $t_{i}$ in the various detectors, given the
actual arrival times $T_{i}$, is given by
\begin{equation}\label{eq:t_i_dist}
  p(t_{i} | T_{i} ) = \prod_{i} \frac{1}{\sqrt{2 \pi} \sigma_{i}}  
  \exp \left[ \frac{- (t_{i} - T_{i})^{2}}{2 \sigma_{i}^{2}} 
    \right] \, ,
\end{equation}
where the timing accuracy $\sigma_{i}$ for each detector is given by
Equation (\ref{eq:sigma_t}), and we assume that timing errors are
Gaussian distributed.

Measurements of the arrival times in each detector can be used to
construct a posterior probability distribution for
the source's sky location $\mathbf{R}$.  This is done by applying Bayes'
theorem to obtain the posterior distribution for the actual
arrival times as a function of the observations as
\begin{equation}\label{eq:post_T_i}
  p(T_{i} | t_{i} ) \propto p(T_{i} )  
  \exp \left[ - \sum_{i} \frac{(t_{i} - T_{i})^{2}}{2 \sigma_{i}^{2}} 
    \right] \, .
\end{equation}
The posterior distribution is the product of the prior distribution
$p(T_{i})$ for the arrival times with the likelihood.
Since we are interested in obtaining a distribution for the sky location
of the event, we would like to re-express (\ref{eq:post_T_i}) in terms
of $\mathbf{R}$.  To do so, we introduce the measured sky position
$\mathbf{r}$ and arrival time $t_{o}$ (in analogy to equation
(\ref{eq:T_i})) as
\begin{equation}\label{eq:t_i}
  t_{i} = t_{o} - \mathbf{r} \cdot \mathbf{d}_{i} \, . 
\end{equation}
Strictly, for an event observed in multiple detectors, the set of
equations (\ref{eq:t_i}) for $t_{o}$ and $\mathbf{r}$ may be
over-determined and not allow any solution.  Although the observed
arrival times of a signal should be consistent with a specific
geocentric arrival time and sky location, in practice, they will not be
precisely consistent due to measurement errors.  In the appendix, we
discuss techniques which can be used to find best fit parameters for
$t_{o}$ and $\mathbf{r}$.  However, for the remainder of this section,
we simply assume that we can eliminate $t_{i}$ in favour of $\mathbf{r}$
and $t_{o}$.

We also make use of (\ref{eq:T_i}) to eliminate $T_{i}$ from
(\ref{eq:post_T_i}) in favour of $\mathbf{R}$ and $T_o$.  The
prior distributions are naturally taken to be uniform over the sphere
(for $\mathbf{R}$) and uniform in time (for $T_{o}$).  Finally, after
marginalizing over $T_{o}$, the posterior distribution for $\mathbf{R}$
is
\begin{equation}\label{eq:posterior}
  p(\mathbf{R} | \mathbf{r} ) \propto p(\mathbf{R}) 
  \exp \left[ - \frac{1}{2} (\mathbf{r} - \mathbf{R})^{T} \mathbf{M} 
  (\mathbf{r} - \mathbf{R}) \right] \, .
\end{equation}
The matrix $\mathbf{M}$, describing the localization accuracy, is
given by
\begin{equation}\label{eq:loc}
  \textbf{M} =  \frac{1}{\sum_{k} \sigma_{k}^{-2}}
  \sum_{i, j} \frac{\mathbf{D}_{ij}
  \mathbf{D}_{ij}^{T}}{2 \sigma_{i}^{2} \sigma_{j}^{2}} \, ,
\end{equation}
where  $\mathbf{D}_{ij} = \mathbf{d}_{i} - \mathbf{d}_{j}$.  A detailed
derivation of this result is provided in the Appendix.  A similar
result has been obtained previously in \cite{Wen:2010cr}.

Equation (\ref{eq:posterior}) provides a simple extension to an
arbitrary number of detectors of the two and three detector result given
in \cite{Fairhurst:2009tc}.  The localization expression has all the
features we would expect, specifically: localization only depends upon
the difference in arrival time between the various detectors;
localization is improved by extending the baseline between detectors and
by better timing accuracy in the detectors; the timing measurement in a
pair of detectors can only serve to restrict the location of the source
in the direction parallel to the detector separation.  The origin of the
normalization pre-factor for the matrix $\mathbf{M}$ arises due to the
marginalization over the geocentric arrival time $T_o$ based upon timing
information at \textit{all} detectors in the network.  

The matrix \textbf{M} is symmetric and can be diagonalized
to obtain three orthogonal eigen-directions $(\hat{e}_x, \hat{e}_y,
\hat{e}_z)$ with localization accuracies $\sigma_{x}, \sigma_{y},
\sigma_{z}$ respectively.  Thus, the posterior distribution for the sky
location is
\begin{equation}\label{eq:prob_xyz}
  p(\mathbf{R} | \mathbf{r} ) \propto p(\mathbf{R})
  \exp \left[ - \frac{1}{2} \left( 
  \frac{\left( x - X \right)^{2}}{\sigma_{x}^{2}} +
  \frac{\left( y - Y \right)^{2}}{\sigma_{y}^{2}} +
  \frac{\left( z - Z \right)^{2}}{\sigma_{z}^{2}} 
  \right) \right] \, ,
\end{equation}
where $\mathbf{R} = (X,Y,Z)$ are the co-ordinates of the source in the
network eigen-basis, and $\mathbf{r} = (x,y,z)$ describes the measured 
location.  The widths $\sigma_{x}, \sigma_{y}, \sigma_{z}$ encapsulate
the ability of the network to localize sources. 

The sky position $\mathbf{R} = (X,Y,Z)$ is restricted to lie on the unit
sphere, and we must take this into account to obtain localization
regions.  Geometrically, equation (\ref{eq:prob_xyz}) describes
ellipsoids of constant likelihood which intersect the unit sphere to
give the localization distribution as an ellipse on the sky.  In most
cases, the source will be localized to a small enough patch of this 
sphere that ignoring its curvature is a good approximation.  In this
case, we project $\mathbf{M}$ onto directions orthogonal to $\mathbf{r}$
using the projection    
\begin{equation}
  \mathbf{P}(\mathbf{r}) = \mathbf{I} - \mathbf{r} \mathbf{r}^{T} \, ,
\end{equation}
where $\mathbf{I}$ is the identity matrix.  This gives
\begin{equation}
  \mathbf{M}(\mathbf{r}) = \mathbf{P}(\mathbf{r}) \; \mathbf{M} \;
  \mathbf{P}(\mathbf{r}) \, .
\end{equation}
The eigenvectors and corresponding eigenvalues of
$\mathbf{M}(\mathbf{r})$ describe the two dimensional localization
ellipse for the source.  The best localization arises when the
projection preserves the two smallest $\sigma$ values, and the worst
when it keeps the two largest.  In all cases, the source is localized
with a probability $p$ within an area
\begin{equation}\label{eq:area} 
  \mathrm{Area}(p) \approx 2 \pi \sigma_{1} \sigma_{2} 
  \left[ - \ln (1 - p) \right] \, ,  
\end{equation}
where $\sigma_{1}$ and $\sigma_{2}$ are the localization accuracies of
the eigen-directions of the projected matrix $\mathbf{M}(\mathbf{r})$. 

When the network is degenerate, one or more of the $\sigma_{i}$ is
infinite and the constant likelihood ellipsoids have infinite extent in
one or more direction.  For a
three detector network, there is one degenerate direction perpendicular
to the plane of the detectors and the surfaces become cylinders with
infinite extent in that direction.  Intersecting the cylinder with the
unit sphere gives two localization regions, one above and the other
below the plane formed by the detectors.  Generically, this degeneracy
can be broken by consideration of the observed amplitudes in the three
detectors, and we assume this can be done.  Taking the normal to the
detector plane to be in the $z$-direction, the localization area for a
three site network can be re-expressed in terms of $\sigma_x$ and
$\sigma_y$ as

\begin{equation}\label{eq:area_3_det} 
  \mathrm{Area}(p) \approx 2 \pi \sigma_{x} \sigma_{y} 
  \left[ - \ln (1 - p) \right]/ \cos{\theta} \, ,
\end{equation}
where $\theta$ is the angle between the z-direction and the source.  If
the source lies in (or close to) the $x-y$ plane, then it cannot be well
localized, and the approximation used to obtain (\ref{eq:area_3_det})
breaks down.  In this case we must go back to the full distribution to
derive the localization region.  For a source lying in the $x$-direction,
we obtain
\begin{equation}\label{eq:area_in_plane} 
  \mathrm{Area}(p) \approx 2 \pi \sigma_{y} \sqrt{2 \sigma_{x}}
  \left[ - \ln (1 - p) \right] \, .
\end{equation}

For a two site network, only $\sigma_{x}$ is finite and the source
location is restricted to a ring on the sky.  The minimal area of the sky
containing the source with a fixed probability $p$ is independent of the
source location and given as:
\begin{equation}\label{eq:area_2_det}
  \mathrm{Area}(p) \approx 
  4\pi \frac{\sqrt{2} \sigma_{x} \mathrm{erf}^{-1}(p)}{D} \, ,
\end{equation}
where $D$ is the separation between the two detectors.

We make repeated use of these localization expressions
(\ref{eq:area}-\ref{eq:area_2_det}) in the remainder of the paper where
we examine the localization capabilities of various detector networks.

%=====================
\section{Advanced detector networks}
\label{sec:adv_networks}
%=====================

Using the results of Section \ref{sec:triangulation}, we can evaluate
the sky coverage and localization provided by various advanced detector
networks.  We take the baseline network to be the \ac{aLIGO} network with two
detectors in Hanford, WA (denoted H) and one in Livingston, LA (denoted
L) supplemented with any number of:

\begin{itemize}
\item the Japanese \ac{LCGT} detector (denoted J) to be located at the
Kamioka mine, 

\item the advanced Virgo detector (denoted V) at Cascina, Italy,

\item the installation of an \ac{aLIGO} detector at the Gingin site in
Australia (denoted A), and consequently only a single \ac{aLIGO} detector at
Hanford.
\end{itemize}

This provides a set of eight advanced detector networks whose
sensitivity and localization ability can be compared.  For simplicity,
we take all advanced detectors to have a \textit{\ac{BNS} horizon} of
360 Mpc and a frequency bandwidth of $100 \mathrm{Hz}$ (corresponding to
a timing accuracy of $0.2 \mathrm{ms}$ at \ac{SNR} 8).  We perform three
different comparison studies.  First, in Section \ref{ssec:fixed_snr},
we consider a source observed in all detectors in the network with an
\ac{SNR} of 8.  Then, in Section \ref{ssec:fixed_dist}, we consider a
set of face on binaries at a fixed distance from various sky positions.
Finally, in Section \ref{ssec:population}, we consider an
astrophysically distributed population of sources.  A subset of these
results were used in presenting the scientific case for installing one
of the advanced LIGO detectors in Australia \cite{LIGOSouth,
AdrianCho08272010}.

%=====================
\subsection{Signal at a fixed \ac{SNR}}
\label{ssec:fixed_snr}

Table \ref{tab:network_localization} summarizes the localization results
for \ac{BNS} sources observed with an \ac{SNR} of 8 in all detectors.
It is rather unlikely that a signal will be observed with the same
\ac{SNR} in all detectors in a network, due to the different antenna
patterns of the detectors.  However, this simple scenario provides a
straightforward method of comparing different networks.  For each of the
eight advanced detector networks, we give the localization widths
$(\sigma_{x}, \sigma_{y}, \sigma_{z})$ from equation (\ref{eq:prob_xyz})
as well as the smallest and largest areas of the 90\% confidence
localization regions.

\begin{table}[t]
\center
\begin{tabular}{| c | c | c | c | }
\hline \hline
Network & Localization(${}^{\circ}$) & Best Area
($\mathrm{deg}^{2}$) & Worst Area ($\mathrm{deg}^{2}$)\\
\hline
\hline
HHL & (1.4, -, -) & 1700 & 1700 \\
AHL & (0.4, 1.7, -) & 8 & 150 \\
\hline
HHJL & (0.5, 2.0, -) & 14 & 220 \\
AHJL & (0.3, 0.7, 4.8) & 4 & 51 \\
\hline
HHLV & (0.5, 1.4, -) & 10 & 150 \\
AHLV & (0.4, 0.6, 1.7) & 3 & 14 \\
\hline
HHJLV & (0.5, 0.5, 2.9) & 4 & 22 \\
AHJLV & (0.3, 0.6, 0.7) & 3 & 6 \\
\hline
\hline
\end{tabular}
\caption{Localization accuracy in networks of detectors.  We assume a
\ac{BNS} signal observed at \ac{SNR} 8 in each detector or,
equivalently, a timing uncertainty of $0.2$ ms in each detector.  The
localization widths correspond to values of $(\sigma_{x}, \sigma_{y},
\sigma_{z})$ from equation (\ref{eq:prob_xyz}) and have been expressed
in degrees.  For the two site HHL network, localization is only possible
in one direction, while for the three site networks there is a
degenerate direction perpendicular to the plane of the network.  The
best and worst case areas are 90\% confidence areas calculated using
equations (\ref{eq:area}-\ref{eq:area_2_det}).}
\label{tab:network_localization} \end{table}

The advanced LIGO network (HHL) comprises only two sites and, as
expected, is unable to provide satisfactory localization for any source.
The three site networks (AHL, HHJL, HHLV) provide good localization in
some directions but degeneracies remain for sources close to the plane
of the detectors.  Although it comprises only three detectors, the AHL
network provides a particularly small ``best case'' localization region
due to the long baselines between the US and Australia.  The results for
the \ac{aLIGO}-Virgo (HHLV) network are somewhat better than
\ac{aLIGO}-\ac{LCGT} (HHJL), even though the baselines are comparable.
The Livingston, Hanford and \ac{LCGT} are close to co-linear, providing
one good localization direction and one relatively poor one.  

The addition of a fourth site to the network (e.g. AHJL, AHLV, HHJLV)
results in a significant
improvement in the ``worst case'' localization.  However, there remain some
directions with relatively poor localization for the AHJL and HHJLV
networks as these sites are nearly co-planar.  The five site network
AHJLV provides good source localization over the full sky.  

%=====================
\subsection{Source at a fixed distance}
\label{ssec:fixed_dist}

\begin{figure}[t]
\centering
\includegraphics[viewport=0 0 500 290, clip, width=.49\textwidth]{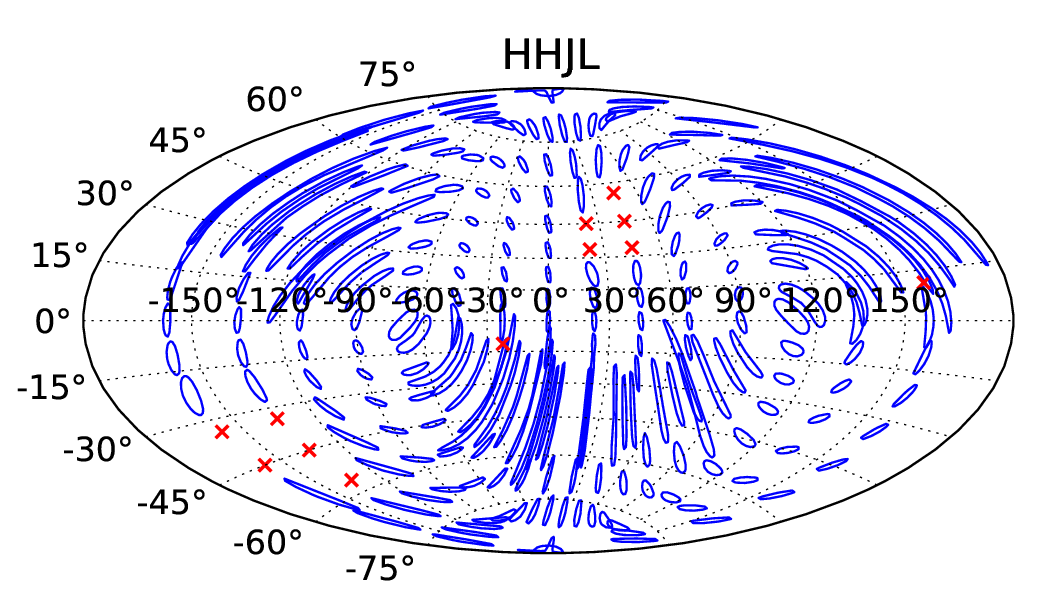} 
\includegraphics[viewport=0 0 500 290, clip, width=.49\textwidth]{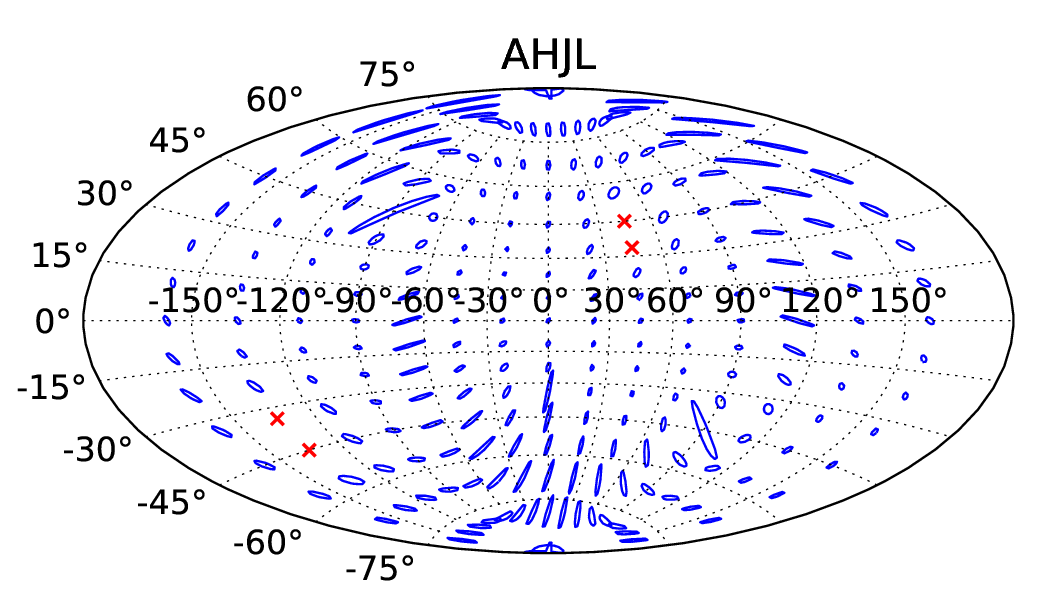}
\includegraphics[viewport=0 0 500 290, clip, width=.49\textwidth]{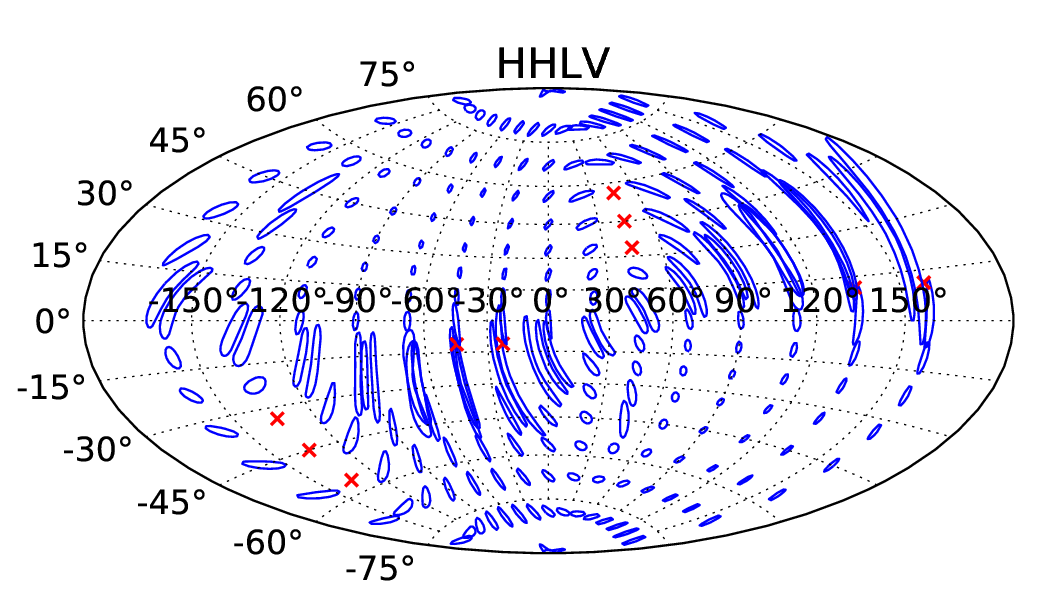}
\includegraphics[viewport=0 0 500 290, clip, width=.49\textwidth]{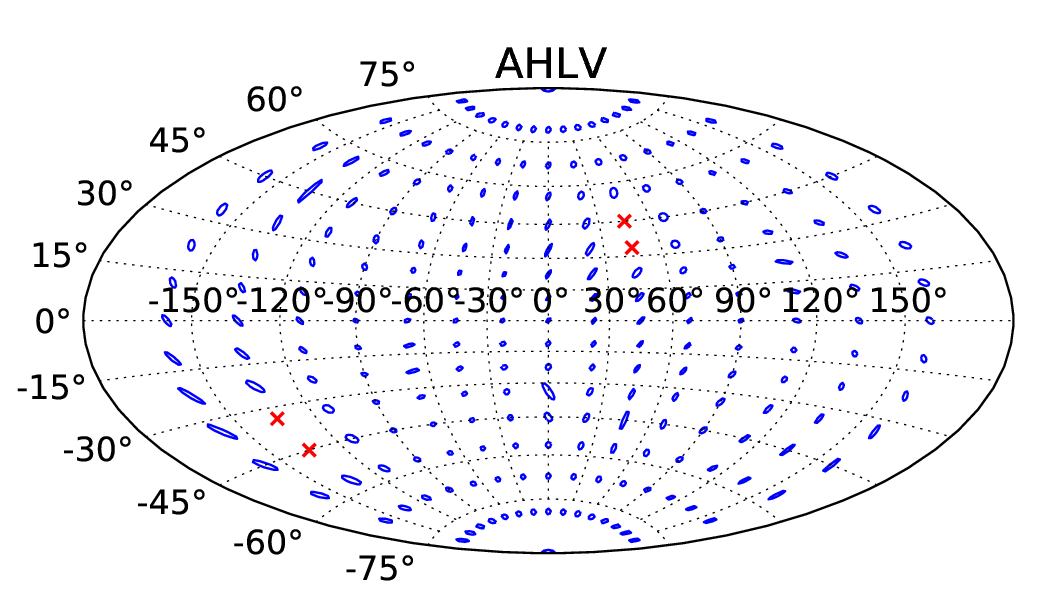}
\includegraphics[viewport=0 0 500 290, clip, width=.49\textwidth]{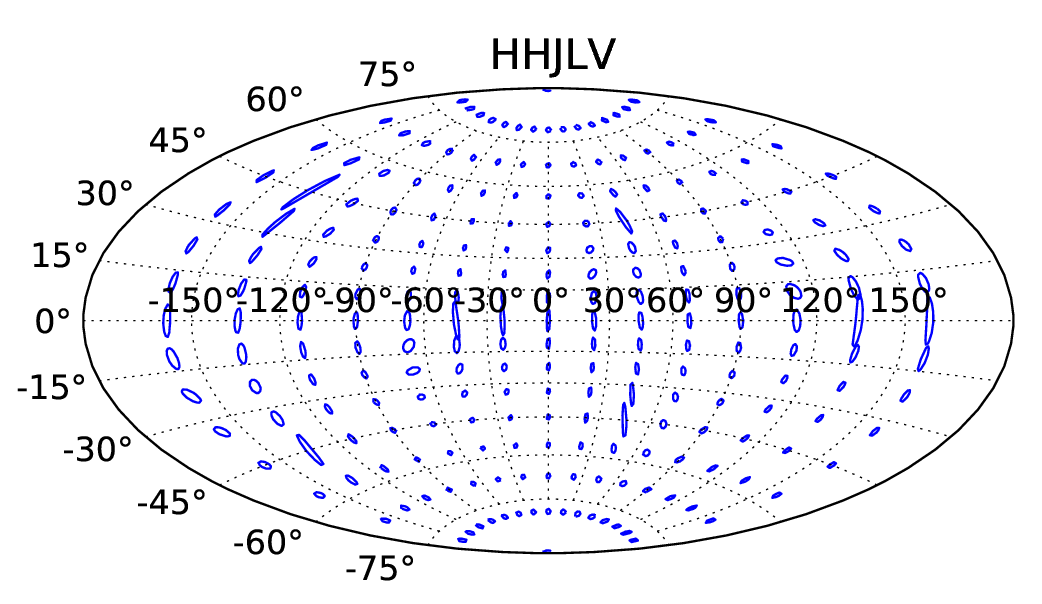}
\includegraphics[viewport=0 0 500 290, clip, width=.49\textwidth]{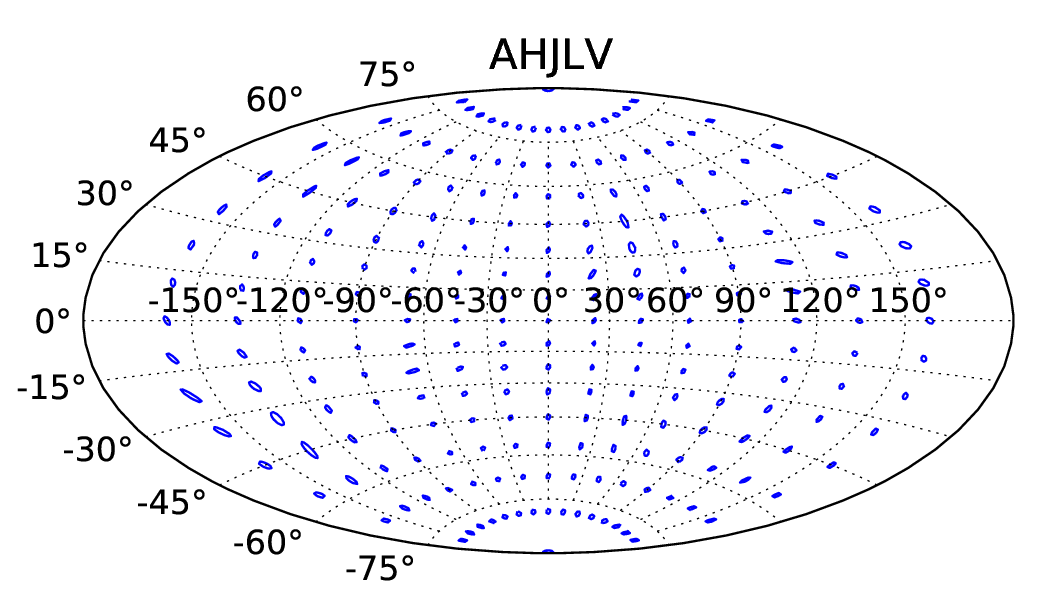}
\caption{The localization accuracy for face on \ac{BNS} at 160 MPc in various
networks of advanced detecors.  The ellipses contain the 90\%
localization regions for sources from varioius points in the sky.  A
$\times$  is plotted at points where the network would not confidently
detect the system.  The plots show the localization for six different
networks: Hanford--Hanford--\ac{LCGT}--Livingston (HHJL);
Australia--Hanford--\ac{LCGT}--Livingston (AHJL);
Hanford--Hanford--Livingston--Virgo (HHLV);
Australia--Hanford--Livingston--Virgo (AHLV);
Hanford--Hanford--\ac{LCGT}--Livingston--Virgo (HHJLV);
Australia--Hanford--\ac{LCGT}--Livingston--Virgo (AHJLV).}
\label{fig:network_localization}
\end{figure}

Here, we consider localization for a face on \ac{BNS} coalescence at a
distance of 160 Mpc at different locations in the sky.  The expected \ac{SNR}
in each detector depends upon its sensitivity to the source direction.
Consequently, we introduce a simple criterion to classify signals as
detectable:  the expected \ac{SNR} in the network must exceed 12, and the
individual \ac{SNR} at two or more sites must exceed 5.  These thresholds are
motivated by results of LIGO-Virgo searches \cite{Abbott:2009tt,
Abbott:2009qj}, where an \ac{SNR} threshold of 5.5 was used in each detector,
and the loudest background events had a combined \ac{SNR} of around 10.  For
those signals which are classified as detectable, the localization
ellipse is calculated using the expressions given in Section
\ref{sec:triangulation}.

Figure \ref{fig:network_localization} shows the localization ellipses
for the four and five detector networks considered in Section
\ref{ssec:fixed_snr}.%
\footnote{Similar results for a different source model have been
presented in \cite{Wen:2010cr}.}
 The results are consistent with those in Section \ref{ssec:fixed_snr}.
A three site network provides good localization in some regions of the
sky, but close to the plane of the detectors the localization ellipses
become extended.  Furthermore, there are regions of the sky where the
signal is deemed undetectable by our criteria.  The addition of a fourth
site breaks the three site degeneracy and significantly improves
localization, particularly for the worst cases.  However, there remain
some slightly extended regions for the AHJL and HHJLV networks which lie
close to the approximate plane formed by the four sites.  The five site
network AHJLV shows excellent resolution over the entire sky.

%=====================
\subsection{A Population of Coalescing Binaries}
\label{ssec:population}

The rate of binary coalescences in spiral galaxies is expected to follow
star formation rate or, equivalently, the blue light luminosity of the
galaxies \cite{Abadie:2010cfa}, and there may well be additional
contributions from elliptical galaxies and globular clusters
\cite{O'Shaughnessy:2009ft}.  However, with detector sensitivities
extending to hundreds of Mpc, it is reasonable to assume a population
which is uniformly distributed in volume.  Additionally, there is no
reason to expect a preferred orientation of binary systems in the
universe, so we take a uniform distribution of binary orientations.  We
use these distributions to simulate the parameters of a large number of
potential signals and for each signal determine whether is is
``detectable'' in a given network, using the same conditions as
previously.  For signals which are detected, we calculate the 90\%
localization area from equations (\ref{eq:area}-\ref{eq:area_2_det}).

\begin{table}[t]
\center
\begin{tabular}{| c | c | c | c | c| c| }
\hline \hline
Network & Detectable Sources & \multicolumn{4}{c|}{Sources Localized within} \\
 & & $1 \deg^2$ & $5 \deg^2$ &$10 \deg^2$ & $20 \deg^2$ \\    
\hline
\hline
HHL & 59 & 0   & 0 &  0 &  0 \\
AHL & 59 & 0.4 & 5 & 13 & 30 \\
\hline
HHJL & 85 & 0.2 &  2 &  5 & 14 \\
AHJL & 85 & 1   & 14 & 36 & 59 \\
\hline
HHLV & 83 & 0.4 &  5 & l3 & 35 \\
AHLV & 84 & 2   & 21 & 48 & 76 \\
\hline
HHJLV & 112 & 2 & 19 & 47 & 77 \\
AHJLV & 114 & 3 & 34 & 84 & 111 \\
\hline
\hline
\end{tabular}
\caption{Sensitivity and localization capability of various different
advanced detector networks to a population of \ac{BNS} signals.  The 
number of signals is normalized so that there are 40 signals observed
in any single detector above \ac{SNR}=8.  This corresponds to the annual
astrophysical rate estimate presented in \cite{Abadie:2010cfa}.  For each
detected signal, we calculate the 90\% localization area and count those
which are localized within 1, 5, 10 and 20 $\deg^{2}$.}
\label{tab:network_population}
\end{table}

Table \ref{tab:network_population} summarizes the results of the simulation.
For each detector network, the expected number of detectable signals as
well as the number localizable within 1, 5, 10 and 20 $\deg^{2}$ is
given.  The numbers are normalized to give 40 signals with \ac{SNR}
greater than 8 in a single detector, in accordance with the
``realistic'' estimate of the annual astrophysical rate
\cite{Abadie:2010cfa}.  However, there is at least an order of magnitude
uncertainty in the rate of \ac{BNS} signals.  Additionally, some
relatively simplistic assumptions have been made for the detection
threshold.  Thus, the results in Table \ref{tab:network_population}
should be taken as illustrative: significant difference between network
performance are meaningful, but the actual values should not be taken
too literally.  

The results for a population of sources again provide a strong case for
the construction of as many detectors at different sites as possible.
As well as an increase in the absolute number of observable sources,
additional detectors greatly increase the fraction of sources which can
be well localized.  For the HHL network no sources will be localized
within $20 \deg^{2}$.  With the introduction of a third site (AHL, HHLV,
HHJL) a significant fraction (20 to 50\%) of sources are localized
within $20 \deg^{2}$, and the loudest signals may be localized within $5
\deg^{2}$.  The addition of a fourth site to the network (AHLV, AHJL,
HHJLV) further improves localization, with the majority of signals
localized within $20 \deg^{2}$, and as many as $20\%$ to within $5
\deg^{2}$.  The five site network provides the most remarkable results,
with virtually all signals localized within $20\deg^{2}$, a third within
$5 \deg^{2}$ and the loudest to within a square degree.  

\begin{figure}[t]
\centering
\includegraphics[width=.49\textwidth]{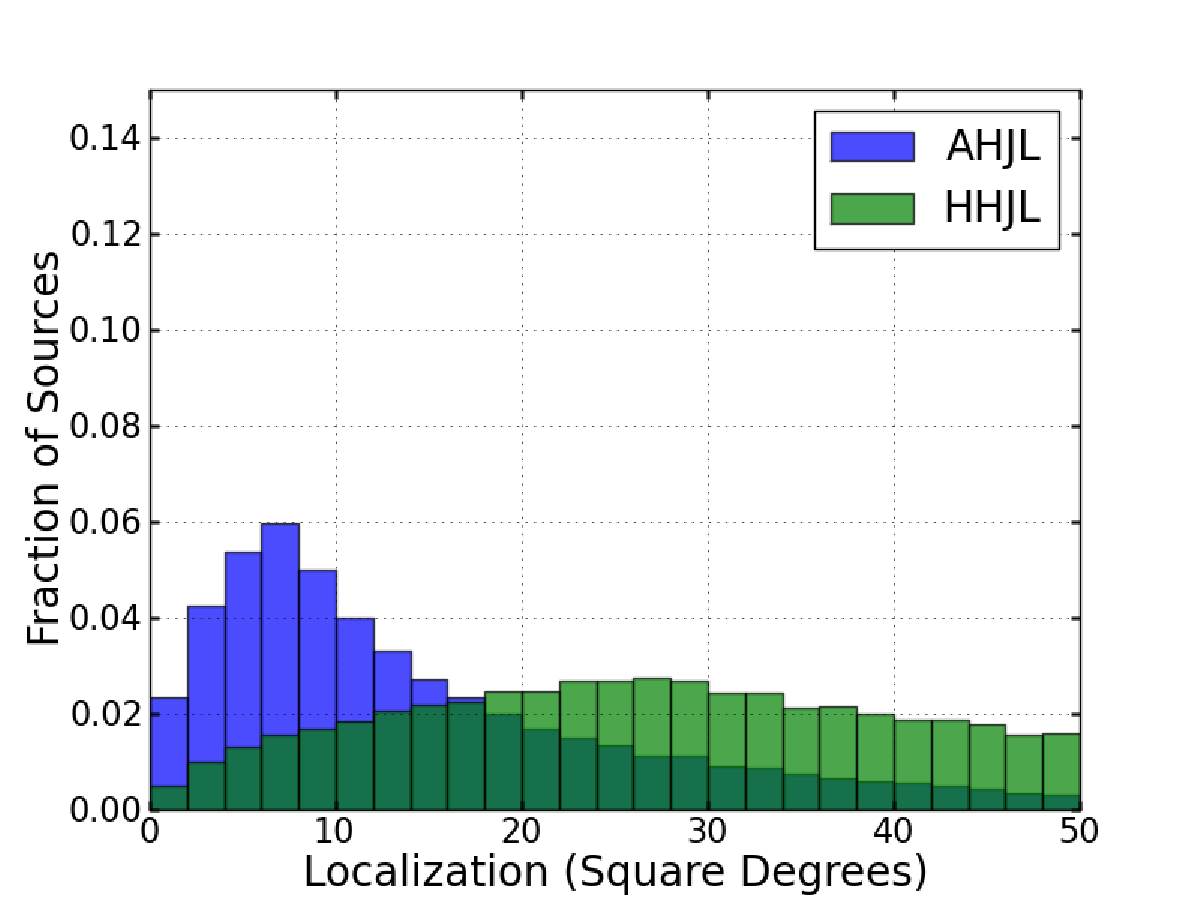}
\includegraphics[width=.49\textwidth]{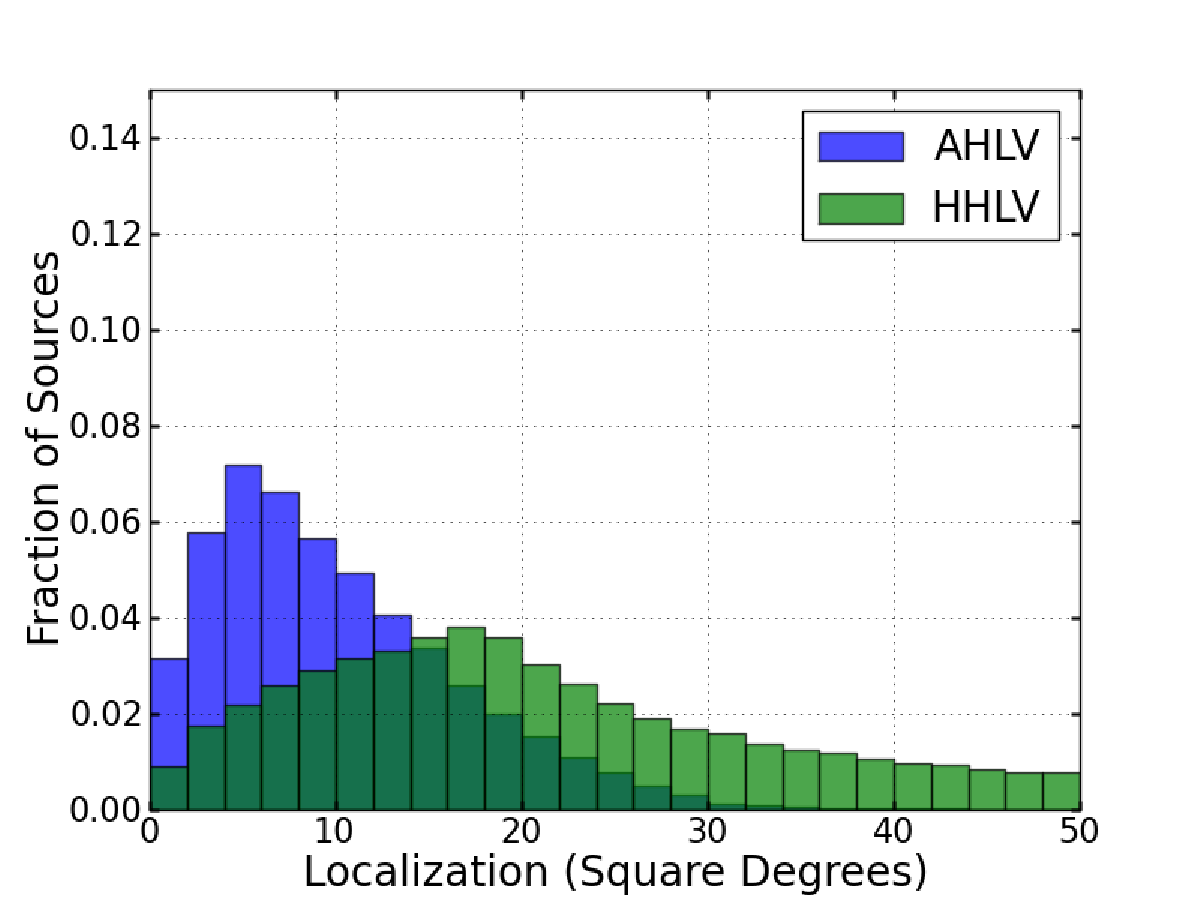}
\includegraphics[width=.49\textwidth]{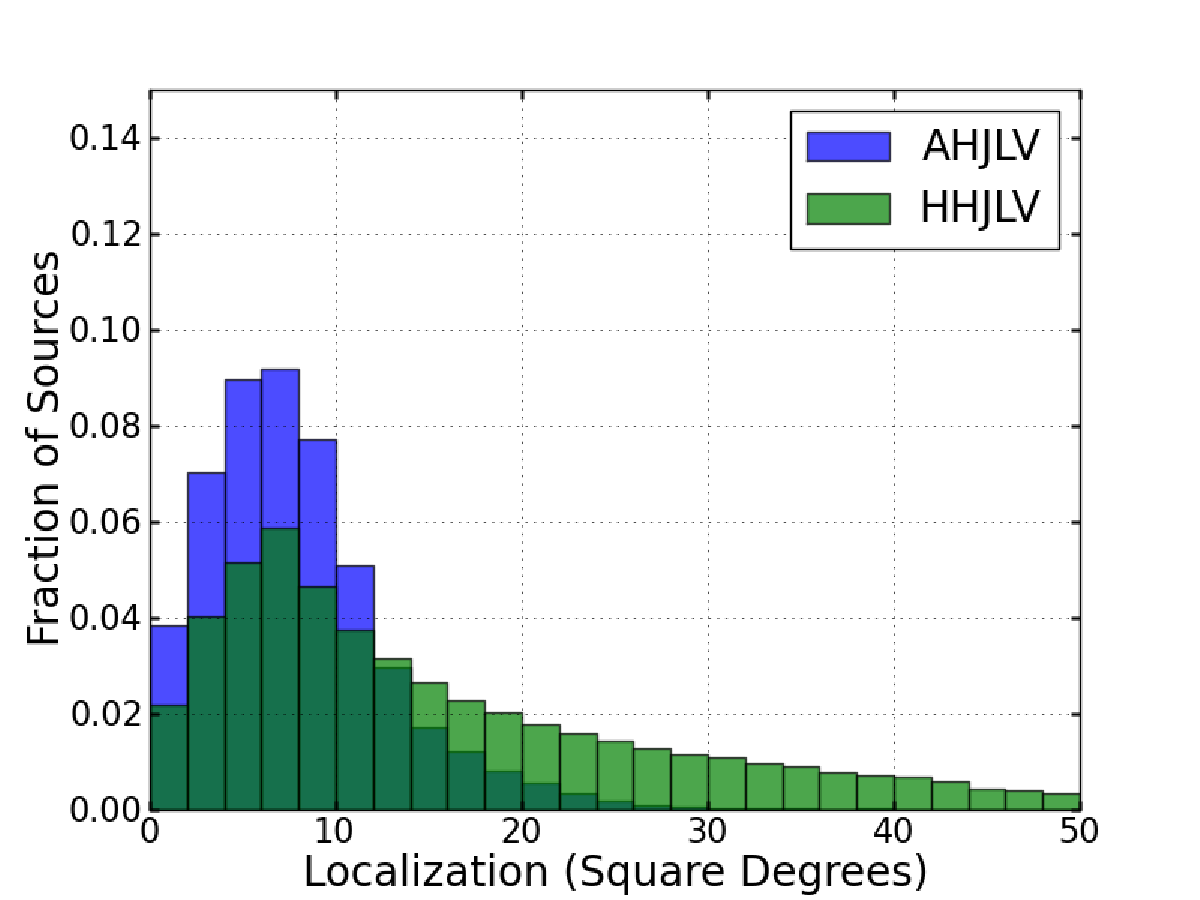}
\caption{Distributions of localization accuracy for various networks} 
\label{fig:network_hist}
\end{figure}

Figure \ref{fig:network_hist} provides a graphical representation of the
same data, showing the localization distribution areas for different
networks.  In all cases, the addition of more sites improves the
localization.  For the networks involving an Australian detector, the
peak of the localization distribution occurs between $5$ and $10
\deg^{2}$.  This corresponds to the typical area of the wide field
electromagnetic transient telescopes currently being operated or under
construction \cite{Rau:2009yx, panstarrs, skymapper, lofar}. 

\section{Discussion}

We have obtained an expression for the localization accuracy of a
gravitational wave signal in a network of detectors.  The localization
expression makes use of only timing information in the various
detectors, with the timing uncertainty taken to be inversely
proportional to both \ac{SNR} and signal bandwidth.  This extends the
results of \cite{Fairhurst:2009tc} to networks with any number of
detectors.  As expected, localization depends only upon the difference
in arrival time between the various detectors, is improved by better
timing accuracy in individual detector and longer baselines between
detectors.  The expressions presented here make numerous simplifying
assumptions by, for example, treating the data as Gaussian and
stationary, using only timing information for localization, taking the
leading order contributions to the timing and localization
distributions, neglecting systematic uncertainties in the waveform and
instrumental calibration.  Thus, while providing useful localization
estimates, a real implementation of source localization would need to
address many additional issues.

We have examined localization of \ac{BNS} sources with a network of
advanced gravitational wave detectors.  It seems reasonable to assume
that the first operational advanced detectors will be at the existing
sites (LIGO Hanford, LIGO Livingston, Virgo) as much of the
infrastructure, including the beam tubes, is already in place.
Furthermore, there is a good chance that early science runs will be
undertaken without a signal recycling mirror (the NOSRM configuration in
Table \ref{tab:bns_time}).  This gives a frequency bandwidth of 40 Hz
for a \ac{BNS} signal, compared to around 100 Hz in the other
configurations.  Consequently, the localization accuracy of the network
could initially be about a factor of five worse than the results
presented in Section \ref{sec:adv_networks}, meaning only the loudest
signals will be well localized.  However, there is the prospect for
significant improvement with the installation of signal recycling to
provide a more broadband spectrum and the addition of new sites (LIGO
Australia, \ac{LCGT}) to the network.  A four site network operating
with broadband sensitivity gives good sensitivity and localization
accuracy over the majority of the sky, with the potential for up to half
of detected \ac{BNS} coalescences to be localized (with $90\%$
confidence) within 10 $\deg^{2}$.  The five site network provides
outstanding coverage over the whole sky with the prospect of virtually
every signal being localized within 20 $\deg^{2}$ and a third within 5
$\deg^{2}$.  These localization areas are commensurate with the field of
view of existing and planned electromagnetic transient observatories
providing the realistic prospect for multi-messenger astronomy in the
advanced detector era.

\label{sec:discussion}

\section*{Acknowledgements}

We have benefited from extensive discussions with many people within the
LIGO Scientific and Virgo Collaborations.  In particular we thank Duncan
Brown, Jolien Creighton, Thomas Dent, Bangalore Sathyaprakash, Bernard
Schutz, Patrick Sutton, John Veitch, Rai Weiss and the LIGO Australia
committee.  We thank Patrick Sutton for independently verifying the
numbers used in Table \ref{tab:bns_time}.  This research was made
possible thanks to support from the Royal Society. 

\appendix

\section{Detailed derivation of the localization expression}

Here, we provide a detailed derivation of the localization expression
(\ref{eq:posterior}) presented in Section \ref{ssec:localization}.  We
begin with equation (\ref{eq:t_i_dist}) which gives the probability
distribution for the signal arrival times $T_{i}$ at the different
detectors, given the measured times $t_{i}$,
\begin{equation}\label{app:post_T_i}
  p(T_{i} | t_{i} ) \propto p(T_{i} )  
  \exp \left[ - \sum_{i} \frac{(t_{i} - T_{i})^{2}}{2 \sigma_{i}^{2}} 
    \right] \, .
\end{equation}
We would like to eliminate the detector arrival times, both measured,
$t_{i}$, and actual, $T_{i}$, in favour of geocentric arrival times and
sky positions.  It is straightforward to replace the actual arrival
times $T_{i}$ with the actual sky location $\mathbf{R}$ (required to be
a unit vector) and geocentric arrival time $T_{o}$ using
\begin{equation}\label{app:T_i}
  T_{i} = T_{o} - \mathbf{R} \cdot \mathbf{d}_{i} \, .
\end{equation}
The observed arrival times will be subject to fluctuations due
to noise in the detectors and the equivalent set of equations for
$t_{o}$ and $\mathbf{r}$ may be over-determined and not allow any
solution.  Typically, a minimization technique is used to solve for the
most consistent sky location and arrival time by minimizing
\cite{Cavalier:2006rz}
\begin{equation}\label{app:error}
  \chi^{2} = \sum_{i} \frac{\left( t_{i} - t_{o} +  \mathbf{r} \cdot
\mathbf{d}_{i}\right)^{2}}{\sigma_{i}^{2}} \, .
\end{equation}
In this appendix, we will not perform the full minimization, but instead
restrict to  the four dimensional surface spanned by $t_{o},\mathbf{r}$,
where we make no restriction on the modulus of $\mathbf{r}$.   Of
course, for a signal, we expect $|\mathbf{r}| \approx 1$.  
We choose not to enforce the exact equality $|\mathbf{r}| = 1$
as it serves to simplify the following derivation.  Working in the
extended parameter space, the minimization can be performed using a
weighted least squares approach.  Differentiating (\ref{app:error})
with respect to $t_{o}$ and $\mathbf{r}$ gives a set of linear
equations
\begin{eqnarray}\label{app:max}
  \sum_{i} \frac{\left(t_{i} - \hat{t}_{o}  +  \hat{\mathbf{r}} \cdot
\mathbf{d}_{i}\right)}{\sigma_{i}^{2}} = 0 \quad \mbox{and} \quad
  \sum_{i} \frac{\mathbf{d}_{i} \left(t_{i} - \hat{t}_{o}  +
  \hat{\mathbf{r}} \cdot \mathbf{d}_{i}\right)}{\sigma_{i}^{2}} = 0 
\end{eqnarray}
which can be solved for the best fit values of the geocentric arrival time
$\hat{t}_{o}$ and ``sky location'' $\hat{\mathbf{r}}$.  These define a
best fit set of arrival times
\begin{equation}\label{app:t_hat_i}
  \hat{t}_{i} = \hat{t}_{o} - \hat{\mathbf{r}} \cdot \mathbf{d}_{i} \, .
\end{equation}
The residual
\begin{equation}
  \chi^{2}_{\mathrm{min}} = \sum_{i} \frac{\left( t_{i} - \hat{t}_{i}\right)^{2}}{\sigma_{i}^{2}}
\end{equation}
encodes the goodness of fit of the timing data to the best fit values.

Having performed this partial maximization, we are in position to
re-express the probability distribution (\ref{app:post_T_i}) in terms of
geocentric arrival times and sky locations.  We begin by expanding 
\begin{eqnarray}\label{app:expanding}
  \sum_{i} \frac{(t_{i} - T_{i})^{2}}{2 \sigma_{i}^{2}}
  &=& \sum_{i} \frac{\left[(t_{i} - \hat{t}_{i})
    + (\hat{t}_{i} - T_{i})\right]^{2}}{2 \sigma_{i}^{2}} \\
  &=& \sum_{i} \frac{\chi^{2}_{\mathrm{min}}}{2} 
    + \frac{1}{2 \sigma_{i}^{2}} \left[ (\hat{t}_{o} - T_{o}) 
    - (\hat{\mathbf{r}} - \mathbf{R}) \cdot \mathbf{d}_{i} \right]^{2}
    \nonumber \\
  &&\quad - \frac{1}{ \sigma_{i}^{2}}\left[t_{i} - \hat{t}_{o} 
    + \hat{\mathbf{r}} \cdot \mathbf{d}_{i} \right] \left[  
    \hat{t}_{o} - T_{o} - (\hat{\mathbf{r}} - \mathbf{R}) \cdot \mathbf{d}_{i}
    \right] \, . 
    \nonumber 
\end{eqnarray}
The first term, encoding the consistency of the measured arrival times,
is constant (independent of $T_{o}$ and $\mathbf{R}$) and will therefore
not affect the localization distribution (\ref{app:post_T_i}).
In addition, the third term vanishes.  This follows directly from the
expressions (\ref{app:max}) used to determine the best fit parameters
$\hat{t}_{o}$ and $\hat{\mathbf{r}}$.  Consequently, the posterior
distribution for sky location and arrival time can be expressed as
\begin{equation}\label{app:post_R_T}
  p(\mathbf{R}, T_{o} | \hat{\mathbf{r}}, \hat{t}_{o}) 
    \propto p(\mathbf{R}, T_{o} )  
    \exp \left[ - \sum_{i} \frac{\left[ (\hat{t}_{o} - T_{o}) 
   - (\hat{\mathbf{r}} - \mathbf{R}) \cdot \mathbf{d}_{i} \right]^{2}}{
   2 \sigma_{i}^{2}} \right] 
\end{equation}

Next, we would like to marginalize over the arrival time $T_{o}$.  The
measurement uncertainties in the arrival time at the various detectors
are typically fractions of a millisecond.  Even in cases where the time
of the event is well known from another astronomical observation, the
arrival time of the gravitational wave signal will \textit{not} be known
with millisecond accuracy.  Consequently, it is appropriate to take a
uniform prior on the arrival time $T_{o}$, and also make the priors on
arrival time and sky location independent.  Then, the distribution
(\ref{app:post_R_T}) can be written as
\begin{eqnarray}\label{app:post_R_T2}
  p(\mathbf{R}, T_{0} | \hat{\mathbf{r}}, \hat{t}_{o}) &\propto& 
    p(\mathbf{R}) \, p( T_{o} ) \, \exp \Bigg[ 
    - (\hat{t}_{o} - T_{o})^{2} \sum_{i} \frac{1}{2 \sigma_{i}^{2}} + \\
  && \qquad 
    2 (\hat{t}_{o} - T_{o})  \sum_{i} \frac{(\hat{\mathbf{r}} - \mathbf{R}) \cdot d_{i}}{ 
    2 \sigma_{i}^{2}}
  - \sum_{i} \frac{[(\hat{\mathbf{r}} - \mathbf{R}) \cdot \mathbf{d}_{i}]^{2}}{
    2 \sigma_{i}^{2}} \Bigg]\, . \nonumber
\end{eqnarray}
In order to marginalize over $T_{o}$, we simply integrate over that
variable.  This is most easily done by completing the square in the
expression (\ref{app:post_R_T2}), and performing the Gaussian integral.
The resulting distribution for the sky location $\mathbf{R}$ is
\begin{eqnarray}
  p(\mathbf{R} | \hat{\mathbf{r}}, t_{o}) &=& \int dT_{o} \, p(\mathbf{R}, T_{0}
| \hat{\mathbf{r}}, \hat{t}_{o}) \\
  &\propto& p(\mathbf{R}) \exp \Bigg[ \left(\frac{-1}{ 2 \sum_{k}
    \sigma_{k}^{-2}} \right) \times \nonumber\\
  &&  \quad \left(
  \sum_{i} \frac{[(\hat{\mathbf{r}} - \mathbf{R}) \cdot \mathbf{d}_{i}]^{2}}{
    \sigma_{i}^{2}} \sum_{j} \sigma_{j}^{-2} - 
    \left[  \sum_{i} \frac{(\hat{\mathbf{r}} - \mathbf{R}) 
    \cdot \mathbf{d}_{i}}{ \sigma_{i}^{2}} \right]^2 \right) \Bigg] \nonumber\\
  &=& p(\mathbf{R}) 
  \exp \left[ - \frac{1}{2} (\hat{\mathbf{r}} - \mathbf{R})^{T} \mathbf{M} 
  (\hat{\mathbf{r}} - \mathbf{R}) \right] \, ,  \nonumber 
\end{eqnarray}
where we have introduced
\begin{equation}\label{app:loc}
  \textbf{M} =  \frac{1}{\sum_{k} \sigma_{k}^{-2}}
  \sum_{i, j} \frac{\mathbf{D}_{ij}
  \mathbf{D}_{ij}^{T}}{2 \sigma_{i}^{2} \sigma_{j}^{2}} 
  \quad  \mathrm{and} \quad
  \mathbf{D}_{ij} = \mathbf{d}_{i} - \mathbf{d}_{j}
\end{equation}
This localization distribution is valid even the location vector
$\hat{\mathbf{r}}$ is not unit.  The restriction that the source
originate from a point on the sky is imposed by the use of a prior
distribution $p(\mathbf{R})$ which enforces $|\mathbf{R}|=1$.  In the
main body of the paper (Equations (\ref{eq:posterior})
and (\ref{eq:loc})), we have made the simplifying assumption that the
measured times $t_{i}$ \textit{are} consistent with a sky location
$\mathbf{r}$, where $|\mathbf{r}| = 1$.  While not strictly necessary,
it does serve to simplify the discussion of localization regions.

\section*{References} 

\bibliographystyle{iopart-num}
\bibliography{refs,iulpapers,ninja}

\end{document}